\documentclass[10pt,a4paper,longbibliography,aps,notitlepage]{revtex4-2}
\usepackage[top=2 cm, bottom=2 cm, left=2 cm, right=2 cm]{geometry}
\usepackage[utf8]{inputenc}
\usepackage{amsmath}
\usepackage{amsfonts}
\usepackage{amssymb}
\usepackage{graphicx}
\usepackage{romannum}
\usepackage{braket}
\usepackage{xcolor}

\begin{document}
\title{Combined time and frequency spectroscopy with engineered dual comb spectrometer}
\author{Sutapa Ghosh} 
\email{Corresponding author}
\author{Gadi Eisenstein}
\affiliation{Andrew and Erna Viterby Department of Electrical Engineering and Russell Berrie Nanotechnology Institute, Technion-Israel Institute of Technology, Haifa 32000, Israel 
\\ \href{mailto: sutapa.g@campus.technion.ac.il}{sutapa.g@campus.technion.ac.il}}

\begin{abstract}
Dual comb spectroscopy (DCS) is a powerful technique for broadband spectroscopy with high precision and fast data acquisition. High-frequency resolution requires long data acquisition times, limiting the temporal resolution in time-resolved measurements. Here we overcome this limitation by engineering the DCS pulse train that interacts with the sample. The measurement is performed in steps where the number of pulses interacting with the sample varies in each step. The DCS spectrum is recorded in each stage, and a multi-dimensional spectrum is generated from which the system time evolution is deduced. We demonstrate this method by measuring the absorption spectrum of a room temperature rubidium vapor. The measured population dynamics of the excited state show a square dependence on the number of interacting pulses due to the coherent accumulation of population. Rabi oscillations are also observed under intense excitation conditions. This is the first demonstration of DCS with high frequency and time resolutions without invoking pump-probe spectroscopy, uniting the pulsed laser spectral and temporal properties. This method can simultaneously detect the kinetics of different chemical species and the pathway for the chemical reaction.
\end{abstract}

\maketitle
\section{Introduction}

Dual-comb spectroscopy (DCS) is a powerful broadband method offering high resolution, fast acquisition times, and mechanical stability as no moving part is involved~\cite{coddington_16,pique_19}. DCS developed hand in hand with advancements in ultra-short pulse technology and laser stabilization methods. One of the comb lasers passes through the sample whose characteristics are imprinted on the laser spectrum. After beating with the second laser, these characteristics are retrieved in the RF domain. Since its first demonstration~\cite{keilmann_04,brehm_06,yasui_05}, the DCS method has been used to characterize a large variety of chemical species~\cite{baumann_11,zolot_13,rieker_14,sajid_14,spearrin_15}.
Nevertheless, even though a pulse laser probes the sample, the dynamical properties of the sample are difficult to extract due to the long data acquisition times needed for high spectral resolutions and signal-to-noise ratios (SNR)~\cite{coddington_16}.

Recently, there have been considerable efforts to combine the time aspects of the pulsed laser in the DCS method to perform time-resolved measurements~\cite{abbas_19,luo_22}. A DCS system combined with an amplitude-modulated CW laser was used to perform time-resolved measurement with a time resolution of 100 $\mu$s, measuring a spectral span of 15 GHz with a frequency resolution of 100 MHz~\cite{huh_21}. Here, the change in the absorption spectrum of the P(27) line of the $\nu_1 + \nu_3$ band in acetylene was measured as a function of an increase in the gas pressure. In another experiment, the time-resolved DCS method showed simultaneous production of $C_2H_6$ and the vibrational excitation of $CH_4$ molecules in a $CH_4$/He gas mixture, in the presence of an electric discharge with a time resolution of 20 $\mu$s~\cite{fleiser_14}. Various techniques have been introduced to increase the time resolution, such as intra-cavity optical filtering that increases the comb tooth spacing, which improves the data acquisition speed and the time resolution $\mu$s~\cite{hoghooghi_21}.

Conventional DCS, in combination with pump-probe spectroscopy, was shown to attain time resolutions of the order of femtoseconds~\cite{asahara_17,kim_20}, set by the delay between the pump and the signal laser, which is controlled by moving a mirror pair on a mechanical translation stage. A translation stage reduces the mechanical stability of the system. Also, its motion is slow, which affects the data acquisition speed. An approach to replace the translation stage in pump-probe DCS spectroscopy has been reported where another comb with a slightly different repetition rate compared to the pump comb was used~\cite{lombsadze_18}. This automatically scans the signal laser in time and later combines with the reference laser to generate the DCS spectrum. The data acquisition speed and stability are preserved at the cost of the complexity involved in using three stabilized comb lasers.

In this paper, we propose and demonstrate a modulated dual-comb spectroscopy method where engineering the DCS pulse train that probes the sample enables high temporal resolutions determined by the inverse laser repetition rate. The interferogram is measured by mixing it with a reference comb laser. The DCS is performed in steps where the number of pulses interacting with the sample varies, forming a multi-dimensional spectrum of the spectral and temporal domains, revealing the sample temporal dynamics while preserving the DCS broadband spectrum. We use the modulated DCS method to measure the multi-dimensional DCS spectrum of room temperature rubidium atoms. The population dynamics of an excited state of various hyperfine transitions of rubidium isotopes were observed. The square dependence of coherent population accumulation of rubidium atoms interacting with the comb laser was measured as a function of the number of interacting pulses. The population dynamics also showed Rabi oscillations under intense excitations. We measured a spectral span of 20 GHz with a frequency resolution of 250 MHz, and the population evolution was measured with a temporal resolution of 4 ns.

\begin{figure}[t]
\begin{center}
\includegraphics[width=\textwidth]{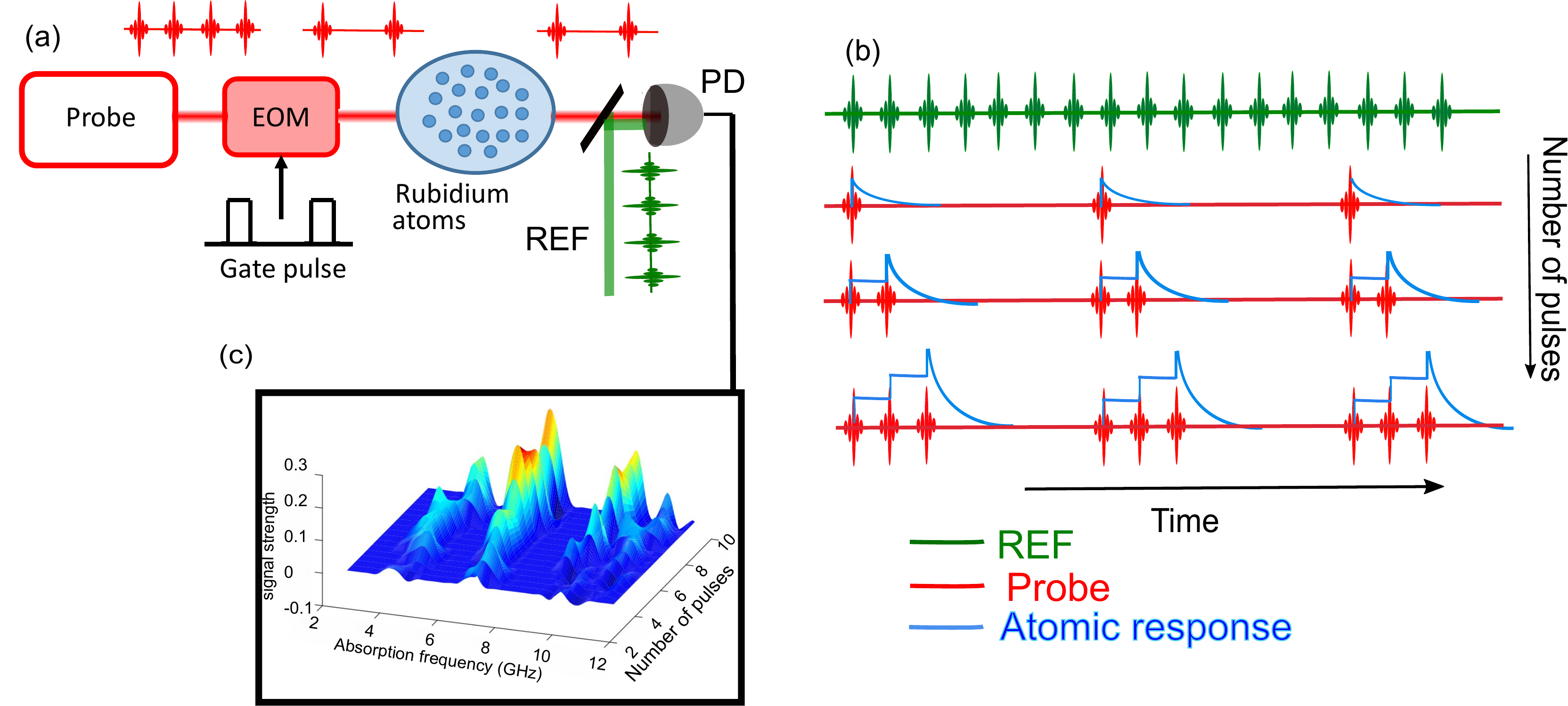}
\caption{{\bf Principle of modulated dual comb spectroscopy} (a) A probe laser (red) generates a pulse train which is modulated by an electro-optic modulator, EOM driven by a rectangular gate pulse. The modulated pulse train passes through the rubidium atoms, mixes with the reference laser, (green) and detected by the photo-detector, PD. (b) Atomic response (blue) as a function of the number of interacting probe pulses (red) which is controlled by the width of the modulating gate pulses.(c) Generated multi-dimensional spectrum where the absorption dynamics of the sample is retrieved as a function of number of interacting probe pulses.}
\label{fig1}
\end{center}
\end{figure}

\section{Results}
{\bf Principle of modulated dual-comb spectroscopy.} The minimum frequency resolution of the DCS spectrometer is given by the repetition rate of the laser probing the sample. To resolve all the lines of the comb, requires a minimum data recoding time of $1/\delta f_{rep}$ where $\delta f_{rep}$ is the difference in repetition rates of the two DCS comb lasers. In practice, the acquisition time can vary from a few hundreds of microseconds to seconds, determined by the spectral width to be measured and the required SNR. All these requirements limit the ability to extract time-resolved properties using the DCS method.

The technique we propose and demonstrate, lifts this limitation by using an engineered pulse train as one of the DCS laser sources. The laser pulse train is amplitude modulated with a rectangular pulse whose width determines the number of pulses that can interact with the sample, as shown in Fig.~\ref{fig1}(a). The modulation repetition rate is chosen, such that the sample fully relaxes between two consecutive modulation cycles. This makes the measured dynamic response repetitive with each modulation cycle, which is then sampled by the reference laser and detected by a photodetector, as shown in Fig.~\ref{fig1}(b). In the following cycle, the modulation pulse widens, allowing more pulses to interact with the sample. Repeating these sequences generates a multidimensional spectrum that combines the frequency and time dependencies of the sample properties. The electric field of the modulated probe pulse train is, $E_{mod} = y(w,t) E_p(t)$, where, $y(w,t)$ is the modulating electric field with a width, $ w$ and $E_p$ represents the probe laser pulse train. The Fourier transform of the modulated pulse train is:
\begin{equation}
E_{mod}(f) = \frac{1}{2\pi T_{rep} T} \sum_l \sum_m \mathcal{F}[y_0](f-lf_{rep}) \mathcal{F}[E_0](lf_{rep}) \delta(f-lf_{rep}-\frac{m}{T})
\end{equation}
The repetition rate of the modulation signal, $1/T$, is chosen to be longer than the atomic relaxation time and shorter than the pulse train repetition rate, $f_{rep}$.  $\mathcal{F}[y_0]$ and $\mathcal{F}[E_0]$ are the Fourier transform of $y(w,t)$ and $E_p(t)$. The absorption spectrum is measured in each step by mixing the two lasers. The data is acquired for a time that ensures a sufficient frequency resolution and a high SNR. The Fourier spectral intensity of the modulated probe laser depends linearly on the number of pulses in each modulation cycle, as shown in section I (Fig. S1) of the supplementary section. The multi-dimensional spectrum represents the absorption spectrum of the sample as a function of the number of pulses probing the sample. To check the validity of our method numerically, we simulated a sample with a quadratic population dependence on time which we were able to retrieve by calculating the  multi-dimensional spectroscopy with the modulated DCS method, as shown in  section II of the supplementary section. 

The sensitivity of the DCS measurement depends on $\delta f_{rep}$ which determines the overlap of the pulses in the time domain. A small $\delta f_{rep}$ leads to a larger number of overlapped pulses, ensuring a high SNR. However, this requires longer data acquisition times which demands a long mutual coherence time of the DCS system.

\begin{figure*}[t]
\begin{center}
\includegraphics[width=\textwidth]{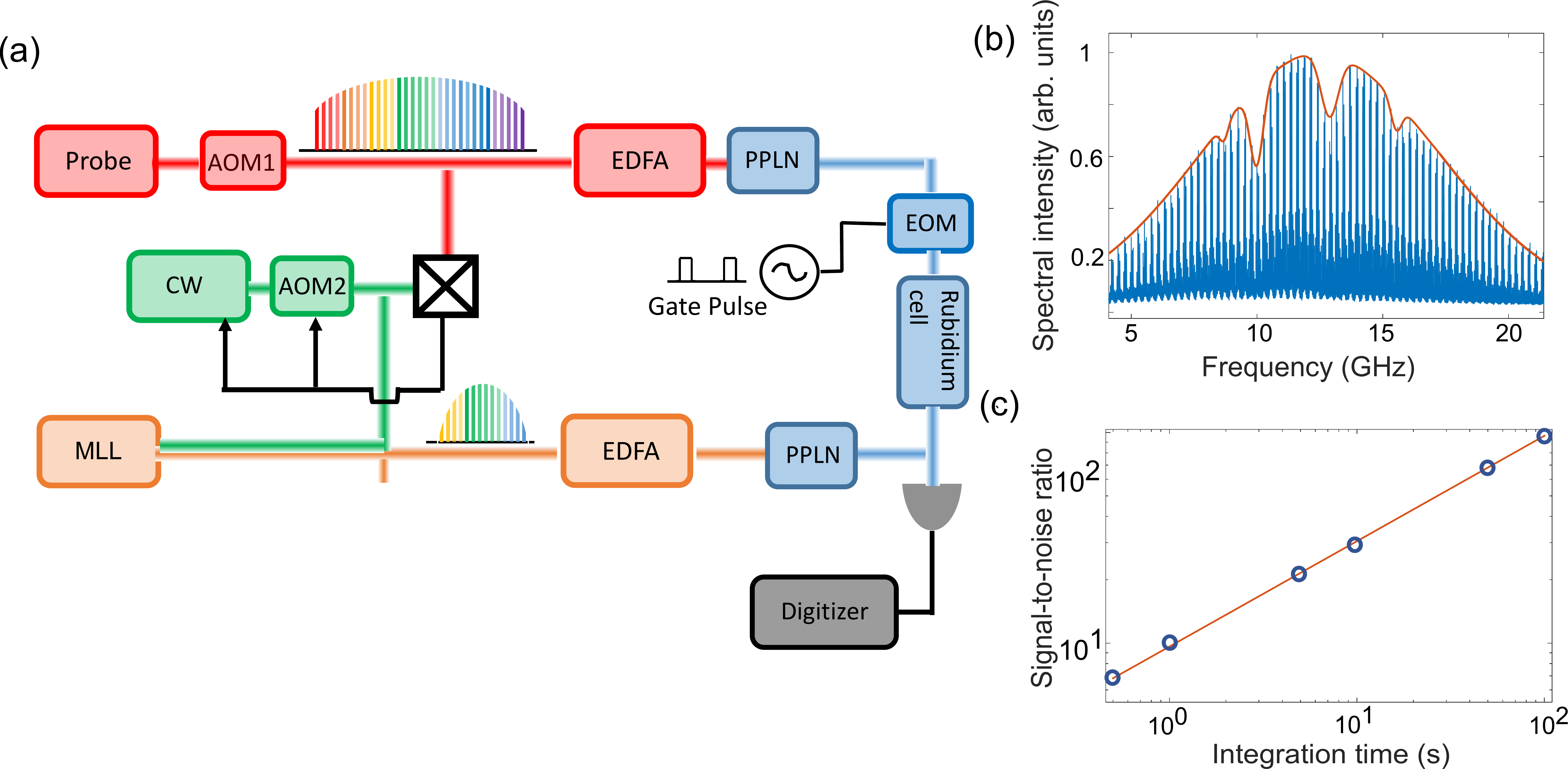}
\caption{{\bf Schematics of the hybrid set-up for dual-comb spectroscopy.} (a) A commercial frequency comb (probe)  is used to stabilize a CW  laser which injection locks the reference laser which is a semi-conductor actively mode locked laser, MLL. 
This transfers the fluctuations of the probe laser to the carrier envelop offset (CEO) of the MLL. The repetition rate of the MLL is separately stabilized against the probe laser by deriving its RF drive signal from the probe laser repetition rate. Both probe laser and MLL are amplified with an erbium-doped fiber amplifier (EDFA) and frequency doubled by periodically poled lithium niobate crystal (PPLN). The probe laser interrogates the rubidium cell and is combined with the MLL to generate the RF beats. (b) The RF domain beat signal shows Doppler-broadened transitions from the mixture of rubidium isotopes, $^{85}$Rb and $^{87}$Rb at room temperature. (c) The square root dependence of signal to noise ratio of the rubidium spectrum with the data integration time, measured for up to 100 seconds. }
\label{fig2}
\end{center}
\end{figure*}

{\bf Demonstration of modulated dual-comb spectroscopy} For the experimental implementation, we use an erbium-doped broadband fiber comb laser with a repetition rate of 250 MHz as the probe. The reference laser is an actively mode-locked semiconductor laser  (MLL) detuned from the probe by $\delta f_{rep}=25$ kHz. Both lasers are amplified, and frequency-doubled. The probe laser passes through an electro-optic amplitude modulator (EOM) whose output is coupled to a room temperature rubidium vapor cell. The experimental setup is shown in Fig.~\ref{fig2}(a). The probe laser is RF locked to a GPS signal that provides absolute long-term stability of $2 \times 10^{-12}$ at 1 second. The reference laser is injection locked by a CW laser which is locked, in turn, on one comb line of the probe laser. A mutual coherence time of 100 seconds was achieved in the present DCS system. This is shown in Fig.~\ref{fig2}(b), which describes the measured DCS spectrum of the rubidium vapor. The SNR  is plotted in Fig.~\ref{fig2}(c) as a function of different integration times; it confirms the square root dependence of SNR on the integration time up to 100 seconds~\cite{ghosh_22}. The EOM is driven by an arbitrary waveform generator (AWG) that applies a rectangular pulse with a repetition rate of $1/T$ and width, $w$ thereby selecting the number of interacting pulses. The pulses interact with the sample, and by the end of each cycle, the sample relaxes to its ground state so that the sequence is repeated at the modulation pulse train repetition rate. The data is recorded for a sufficiently long time to ensure the required spectral resolution. In the present experiment, $\delta f_{rep}$ was $25$ kHz, meaning that a single interferogram is $1/\delta f_{rep} = 40 \mu$s long. Since the reference laser pulse width is around $50$ ps, the number of overlapped pulses was  $\Delta t_{width} f_{rep}^2/\delta f_{rep} = $ 100 in each interferogram. For a modulation repetition rate of 25 MHz, the number of overlapping pulses for the minimum modulation width decreases to 10. The data acquisition time in the current experiment was $50$ ms, yielding 1250 interferograms that were averaged and Fourier transformed. The rubidium DCS spectrum was measured as a function of the number of pulses that interacted with the rubidium atoms.

\begin{figure*}[t]
\begin{center}
\includegraphics[width=\textwidth]{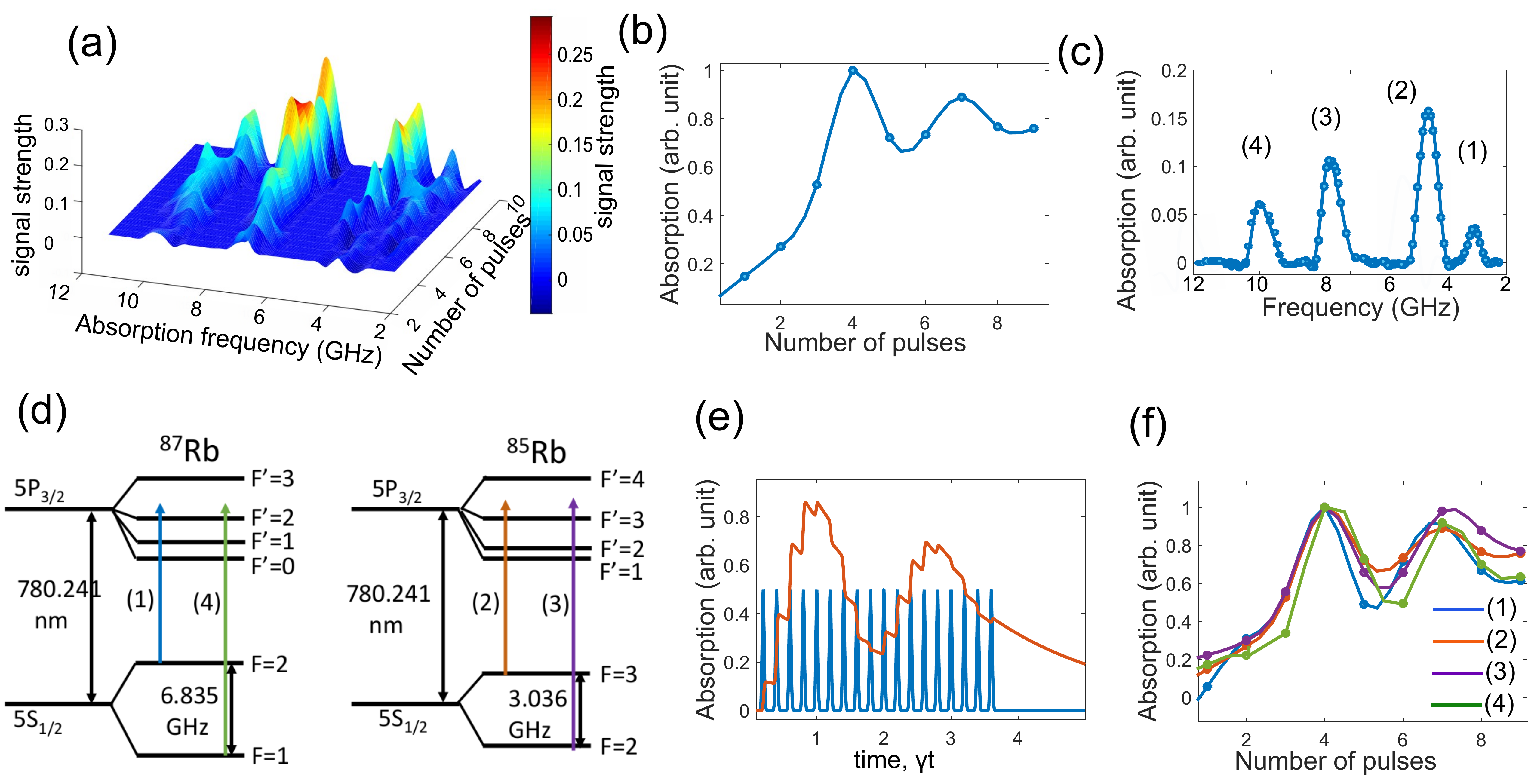}
\caption{{\bf Multi-dimensional dual comb spectroscopy.} (a) Experimental results of the multi-dimensional spectrum as a function of the absorption frequency and the number of interacting pulses. (b) The vertical section represents the population dynamics probed with each pulse cycle of the probe laser. (c) The horizontal section represents the absorption spectrum of the Doppler broadened transitions of the rubidium isotopes. (d) The schematics of level diagram for the rubidium isotopes $^{85}$ Rb and $^{87}$Rb are marked. (e) Simulation results of the population evolution of the excited state (red) interacting with the pulse train (blue). (f) Temporal evolution of population corresponding to each transition of the rubidium mixture.}
\label{fig3}
\end{center}
\end{figure*}

{\bf Multi-dimensional spectroscopy of rubidium atoms.} We study a two-level atoms (rubidium vapor) excited by a train of pulses. The Hamiltonian for this system is $H = H_{atom} + H_{int}$, where $H_{atom} = \hbar \omega_0 \sigma^\dagger \sigma $ and $H_{int} = \hbar \Omega (\sigma e^{i\omega t} + \sigma^{\dagger} e^{-i\omega t}) $, $\sigma = \ket{g}\bra{e}$ and $\Omega$ is the Rabi frequency $\Omega = -\frac{\braket{g|(\hat{\epsilon}.d)|e} E_0(t)}{\hbar}$, $d$ being the dipole moment of the atom interacting with the pulse train electric field, $E_0(t)$.

 Since the inter-pulse separation is smaller than the atomic relaxation times, the population of atoms is coherently accumulated with each pulse and shows a square dependence on the number of accumulated pulses~\cite{felinto_03,marian_04}. If the excitation is sufficiently strong, the population transfers periodically between the ground and excited states exhibiting Rabi oscillations~\cite{huber_11}. Usually, it is difficult to observe Rabi oscillations in the atomic ensemble at room temperature due to fast de-coherence times. Probing atoms with a high-intensity pulse train can drive fast coherent dynamics in the system, which can then be measured using the modulated DCS method. The rubidium atoms were excited with a modulated pulse train containing different numbers of pulses, each separated by 4 ns. The relaxation time, of both rubidium isotopes $^{85}$Rb and $^{87}$Rb is around 27 ns for the $5P_{3/2}$ state. In the experiment, we recorded the absorption spectrum corresponding to various hyperfine transitions. The absorption spectrum as a function of the number of interacting pulses is plotted as a multi-dimensional spectrum in Fig.~\ref{fig3}(a). The vertical section is plotted in Fig.~\ref{fig3}(b). It represents the temporal evolution of the population in the excited state of the $5S_{1/2} \rightarrow 5P_{3/2}$ transition in a $^{85}$Rb isotope. The square behavior of the population transfer represents the coherent accumulation of the pulses in the system, while the oscillation represents the Rabi oscillation at $2\pi \times 2.113$ MHz. The horizontal section is shown in Fig~\ref{fig3}(c), representing the rubidium absorption spectrum corresponding to a specific modulation width. The Doppler broadened transitions are marked according to the level diagram shown in Fig.~\ref{fig3}(d). Fig.~\ref{fig3}(e) shows simulation results of the Bloch equation corresponding to the Hamiltonian for the parameters of the present experiment which qualitatively agrees with the data. Fig~\ref{fig3}(f) shows the temporal evolution of all the marked transitions. Since the dipole moment of both the rubidium isotopes for the transition, $5S_{1/2} \rightarrow 5P_{3/2}$ are the same, they exhibit Rabi oscillations with the same frequency~\cite{steck_03}. 

\begin{figure*}[t]
\begin{center}
\includegraphics[width=\textwidth]{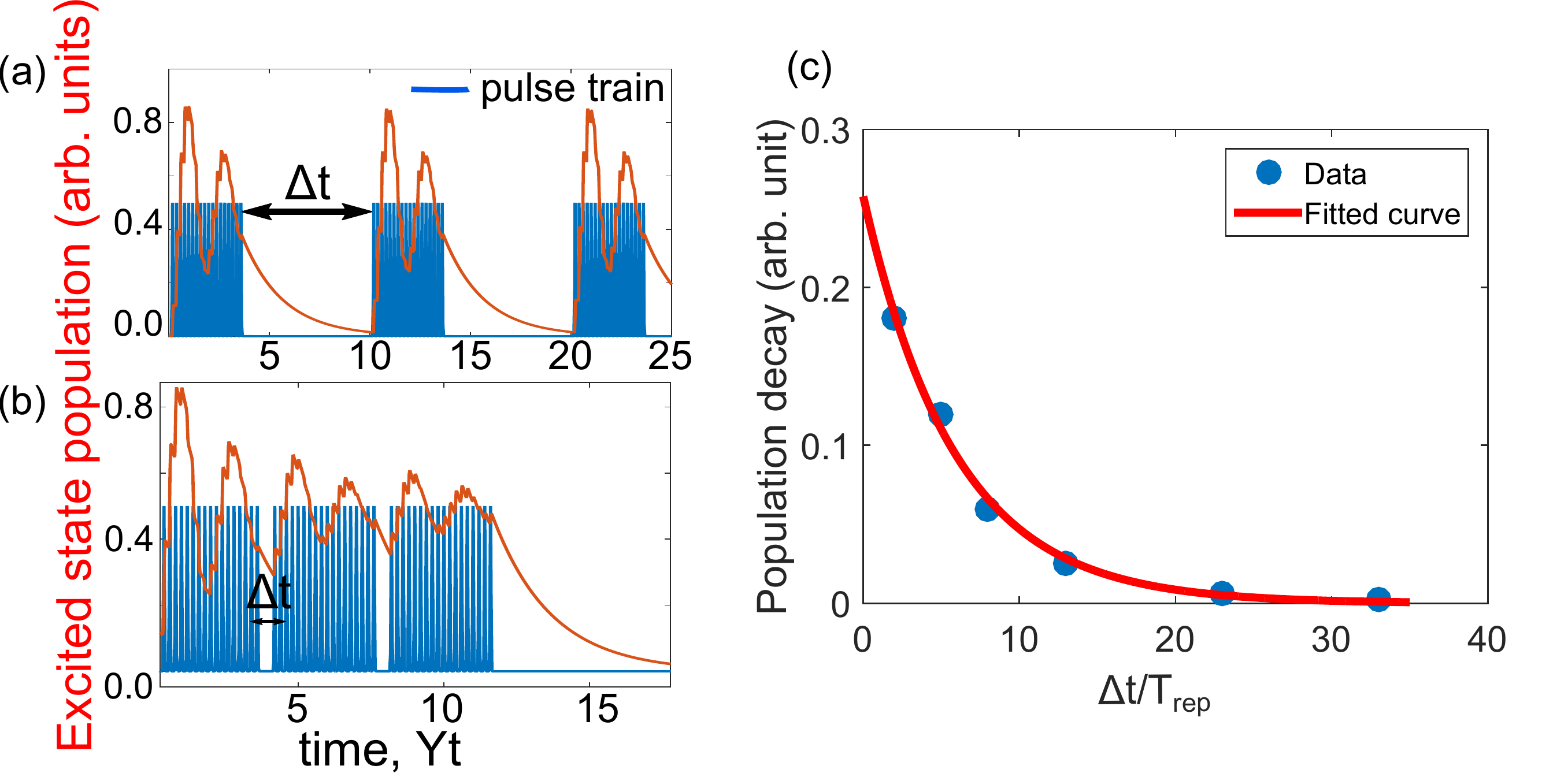}
\caption{{\bf Effect of the atomic relaxation time.} (a) Population evolution with each modulation cycle for the case when the atoms relax to the ground state before each cycle. (b) The amplitude of the Rabi oscillations decreases exponentially for the cases where the atoms do not relaxes completely to the ground state. (c) The population decay constant is plotted as a function of $\Delta t$ .}
\label{fig4}
\end{center}
\end{figure*}

In cases when the modulation repetition rate is such that the atoms are not entirely relaxed to the ground state, the population transfer in each modulation sequence experiences an exponential decay that depends on the system de-coherence rate ($\gamma$). For a high de-coherence rate, the populations decay rapidly to the ground state. Fig.~\ref{fig4}(a) relates to a case where $\Delta t$ is larger than the atomic relaxation rate, and the population transfer is identical for each cycle. In contrast,  Fig.~\ref{fig4}(b), shows the case when the atom do not relax completely, so the strength of the Rabi oscillation decays with each cycle. The simulated result of the oscillation decay constant as a function of different time gap between the modulating sequences, $\Delta t$ is shown in Fig.~\ref{fig4}(c), which describes the exponential decay and depends on the relaxation time of the atoms.  

\section{Discussion}
We introduced a new DCS scheme where the laser pulse train interacting with the sample is engineered. As proof of concept, we applied this method to a gas mixture of rubidium isotopes. We demonstrated that the technique enables broadband spectroscopy separating the various hyperfine transitions of Doppler-broadened rubidium isotopes, and measuring the population evolution into the excited state. Our measurement also shows Rabi oscillation imprinted on the population dynamics. 

The time resolution of the modulated DCS method depends on the laser repetition rate with high repetition rate allow measuring fast processes. In the present set up, the temporal resolution is 4 ns. It is possible of course to use very high repetition rates. An extreme example is a chip scale comb with a repetition rate of 450 GHz~\cite{avik_18} that offers a time accuracy of 2.22 ps, although this also limits the frequency resolution to 450 GHz. However, the relevant transition can still be excited by scanning the comb laser frequency around the molecular excitation frequency by tuning the repetition rate.
   
Combining the time and frequency aspects of a pulsed laser in DCS makes it a powerful spectroscopic tool. This type of spectroscopy is especially beneficial for studying complicated chemical reactions that lead to many metastable and short-lived chemical compounds produced during the reaction~\cite{fleisher_14}. Due to the broadband nature, most chemical species are likely to be detected. At the same time, the temporal aspects can extract information about the path of the chemical reaction from the parent compound mixture to the final product mixture. 

\section{Materials and methods}
Fig.~\ref{fig2} describes an overview of the hybrid dual comb spectroscopy apparatus. The probe laser is a commercial fiber laser (Menlo systems, FC1500-250-ULN) with a repetition rate, $f_{rep}$ of 250 MHz. The reference laser  is a MLL detuned by 25 kHz from the probe laser repetition rate. It employs a piezo-mounted mirror in the cat-eye configuration and an intracavity interference filter. The MLL is injection locked by an external cavity CW laser which is locked, in turn, on a single tooth of the probe laser. The beat between the CW laser and the probe laser is used to stabilize the CW laser by applying a fast-feedback to an acousto-optic modulator, AOM2 and a slow feedback to the piezo mounted cavity mirror. The fully stabilized MLL emits pulses with a duration of 50 ps. 

The probe laser is amplitude modulated by an electro optic Mach-Zehnder modulator from EO space which is driven by an arbitrary waveform generator (AWG), Agilent Technologies, 81150A. To modulate probe laser pulses, the modulation signal from AWG needs to be synchronized to the probe pulses. To this end, we derive an RF signal of 250 MHz from the probe laser repetition rate and used it as a reference to a phase locked oscillator that generates the clock signal for the AWG. a signal with chosen modulating frequency (25 MHz) as a clock signal to the AWG.

Both lasers were tuned to 1560 nm and are amplified by an erbium doped fiber amplifier, EDFA before being frequency doubled using temperature stabilized periodically poled lithium niobate crystals. The frequency doubled modulated probe pulses passes through a vapor cell filled with rubidium isotopes ($^{85}$Rb and $^{87}$Rb) held at room temperature. After interacting with the gas sample in a single-pass configuration, the two frequency-doubled lasers are mixed and detected by a photo-detector, PD, FPD610-FC-VIS. The detected signal is filtered, amplified and digitized. A fast Fourier transform is computed with the background subtracted from the data. The amplitude of the spectrum is retrieved and displayed in Fig.\ref{fig2}(b). 

The numerical simulation to calculate the population dynamics of atoms has been performed with a "Quantum optics toolbox" in Julia with a modifications to introduce the effect of pulse train~\cite{kramer_18}.

\def\bibsection{\subsection*{\refname}} 
\bibliography{mod_DCS}

\end{document}